\documentclass[US Letter]{article}
\usepackage{spconf,amsmath,graphicx,amssymb,delarray,subfigure,verbatim,booktabs}
\usepackage{diagbox, makecell}
\usepackage{multirow}
\usepackage[english]{babel}
\usepackage{lipsum}

\title{ICASSP 2021 DEEP NOISE SUPPRESSION CHALLENGE: DECOUPLING MAGNITUDE AND PHASE OPTIMIZATION WITH A TWO-STAGE DEEP NETWORK}
\name{Andong Li$^{\star \dagger}$, Wenzhe Liu$^{\star \dagger}$, Xiaoxue Luo$^{\star \dagger}$, Chengshi Zheng$^{\star \dagger}$, Xiaodong Li$^{\star \dagger}$}
\address{$^{\star}$ Key Laboratory of Noise and Vibration Research, Institute of Acoustics, Chinese Academy\\
	of Sciences, Beijing, China\\
	$^{\dagger}$ University of Chinese Academy of Sciences, Beijing, China}
\begin{document}
\ninept

\maketitle
\newcommand\blfootnote[1]{%
	\begingroup
	\renewcommand\thefootnote{}\footnote{#1}%
	\addtocounter{footnote}{-1}%
	\endgroup
}

\begin{abstract}
It remains a tough challenge to recover the speech signals contaminated by various noises under real acoustic environments. To this end, we propose a novel system for denoising in the complicated applications, which is mainly comprised of two pipelines, namely a two-stage network and a post-processing module. The first pipeline is proposed to decouple the optimization problem $w.r.t.$ magnitude and phase, $\emph{i.e.}$, only the magnitude is estimated in the first stage and both of them are further refined in the second stage. The second pipeline aims to further suppress the remaining unnatural distorted noise, which is demonstrated to sufficiently improve the subjective quality. In the ICASSP 2021 Deep Noise Suppression (DNS) Challenge, our submitted system ranked top-1 for the real-time track 1 in terms of Mean Opinion Score (MOS) with ITU-T P.808 framework.

\end{abstract}
\vspace{-0.3cm}
\section{Introduction}
\label{sec:intro}
\vspace{-0.2cm}
In real scenarios, environmental noise and room reverberation may have a negative impact on the performance of automatic speech recognition (ASR) systems, video/audio communication, and hearing assistant devices. To tackle these problems, many speech enhancement (SE) algorithms have already been proposed to effectively estimate the clean speech while sufficiently suppress the noise components~{\cite{loizou2013speech}}. Recent years have witnessed the rapid development of deep neural networks (DNNs) toward SE research~{\cite{xu2014regression, wang2018supervised}}. By embracing a data-driven paradigm, the SE task can be formulated as a supervised learning problem, where the network attempts to uncover the complicated nonlinear relationship between noisy features and clean targets in the time-frequency (T-F) domain.

In the previous studies, only the recovery of magnitude is explored while noisy phase is directly incorporated for speech waveform reconstruction~{\cite{xu2014regression, wang2018supervised}}. The reasons can be attributed to two-fold. For one thing, the phase is considered difficult to estimate due to its unclear structure. For another, previous literature reported that the recovery of phase did not bring a notable improvement in speech perception quality~{\cite{wang1982unimportance}}. More recently, the importance of phase receives continuously increasing attention in improving speech quality and intelligibility~{\cite{paliwal2011importance}}. Williamson $et.al.$~{\cite{williamson2017time}} proposed the complex ratio mask (CRM), where the mask was applied to both real and imaginary (RI) components, and both magnitude and phase could be perfectly estimated theoretically. Afterward, the complex spectral mapping technique was proposed, and the network was required to directly estimate the RI spectrum, which was reported to achieve improved speech quality than masking-based methods~{\cite{tan2019learning}}. More recently, time-domain based methods begin to thrive, where the raw waveform serves as both input and output~{\cite{defossez2020real}}. In this way, explicit phase estimation problem is effectively avoided. Although both classes can obtain impressive performance in objective metrics, we resort to the complex domain method as we found that complex domain based methods achieved overall better mean opinion scores (MOS) than time-domain methods in the INTERSPEECH 2020 Deep Noise Suppression (DNS) Challenge{\footnote{https://dns-challenge.azurewebsites.net/phase1results/Interspeech2020}}. We attribute the reason to better distinguishment for speech and noise in the T-F domain than raw waveform format.

To handle the noise reduction problem in more challenging acoustic environments in the ICASSP 2021 DNS Challenge~{\cite{reddy2020icassp}}, we propose a novel SE system, called $\textbf{T}$wo-$\textbf{S}$tage $\textbf{C}$omplex $\textbf{N}$etwork with a low-complexity $\textbf{P}$ost-$\textbf{P}$rocessing scheme (dubbed TSCN-PP). It mainly consists of two processing pipelines. Firstly, a novel two-stage network paradigm is designed, which is comprised of two sub-networks, namely $\emph{coarse magnitude estimation}$ $\emph{network}$ (dubbed CME-Net) and $\emph{complex spectrum refine network}$ (dubbed CSR-Net). For CME-Net, it offers a coarse estimation toward the spectral magnitude, which is then coupled with noisy phase to obtain a coarse complex spectrum. Afterward, CSR-Net attempts to refine the complex spectrum by receiving both coarse estimation and noisy spectrum as the input. Note that the role of CSR-Net is two-fold. Firstly, instead of directly estimating the spectrum of clean target, it only captures the residual details, $\emph{i.e.}$, the estimated details are added by the input to obtain the final refined spectrum. Secondly, as some noise components still remain, CSR-Net helps to further suppress the residual noise. For the second pipeline, we propose a low-complexity post-processing (PP) module to further alleviate the unnatural residual noise, which is validated to be a significant step in improving the subjective speech quality.

We explain the design rationale of our algorithm from two aspects. Firstly,
the single-stage network often fails to function well for relatively difficult tasks due to its limited mapping capability. More recently, the literature~{\cite{yu2019gradual, li2020speech, li2020recursive}} has revealed the advantage of multi-stage training over single-stage methods in many tasks, $\emph{e.g.}$, image deraining and speech separation. Secondly, due to the nonlinear characteristics of DNNs, some nonlinear distortion may be introduced when the test set mismatches the training conditions. For example, as the SE model is often trained with a wide range of synthetic noisy-clean pairs, when the trained model is applied under a more complicated real environment, it may introduce some unpleasant nonlinear distortion, which substantially degrades the subjective quality. Therefore, it is of necessity to take a PP module to further suppress the residual noise if audible speech distortion can be avoided. In our subjective experiment, we do find that, after applying the PP, the whole subjective quality can be consistently improved.

The remainder of the paper is structured as follows. In Section~{\ref{sec:proposed-tscn-dspp}}, both the proposed two-stage network and the post-processing module are presented. In Section~{\ref{sec:experiments}}, the experimental settings are given. The experimental results are revealed in Section~{\ref{sec:results-and-analysis}}. We draw some conclusions in Section~{\ref{sec:conclusion}}.
\vspace{-0.2cm}
\section{PROPOSED TSCN-PP}
\vspace{-0.2cm}
\label{sec:proposed-tscn-dspp}
\subsection{Notations}
\label{notations}
\vspace{-0.2cm}
We present the diagram of the proposed approach, as shown in Fig.~{\ref{fig:architecture}}. In this paper, we denote the $\left( X_r, X_i \right)$ as the noisy complex spectrum, while $\left(\tilde S^{cm}_{r|i}, S_{r|i} \right)$, $\left(\tilde S^{cs}_{r|i}, S_{r|i} \right)$, and $\left(\tilde S^{pp}_{r|i}, S_{r|i} \right)$ are the real and imaginary parts of the estimated and the clean speech after CME-Net, CSR-Net and PP module, respectively. In addition, the mapping functions of CME-Net, CSR-Net and PP-Module are defined as $\mathcal{F}_{cm}$, $\mathcal{F}_{cs}$, and $\mathcal{F}_{pp}$ with the parameter sets being ${\bf \Phi}_{cm}$, ${\bf \Phi}_{cs}$, and ${\bf \Phi}_{pp}$, respectively.
\vspace{-0.2cm}
\subsection{Two-stage Network}
\label{two-stage-network}
\begin{figure*}[t]
	\centering
	\centerline{\includegraphics[width=1.45\columnwidth]{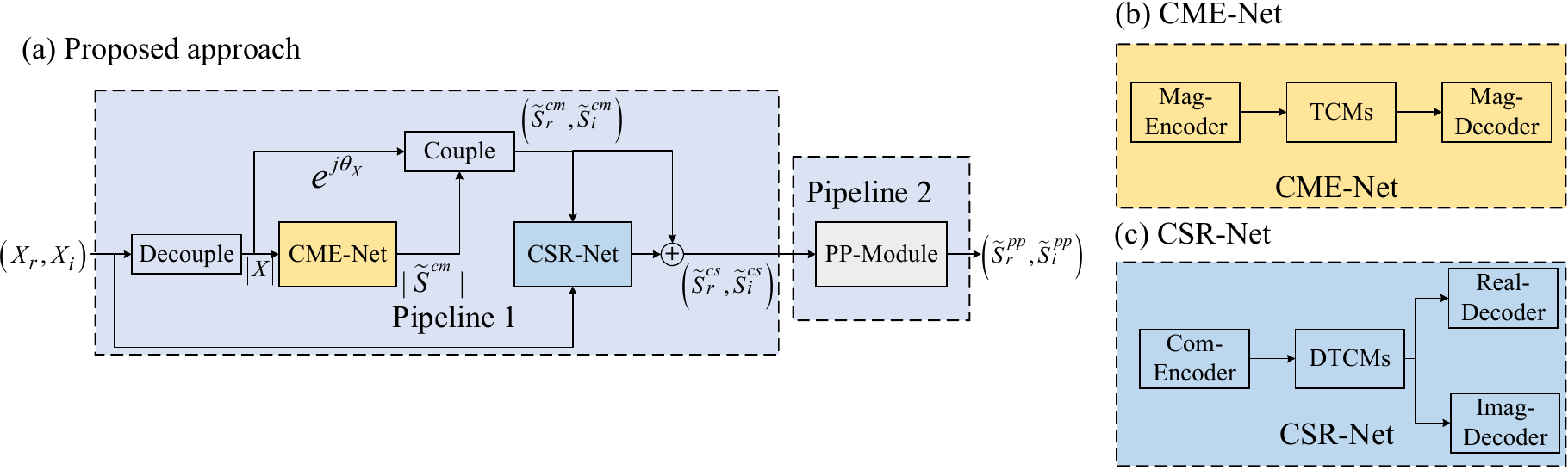}}
	\caption{Proposed processing pipelines for the ICASSP 2021 DNS Challenge. (a) Proposed two-stage framework with postprocessing. (b) The network detail of CME-Net. (c) The network detail of CSR-Net.}
	\label{fig:architecture}
	\vspace{-0.4cm}
\end{figure*}

\vspace{-0.2cm}
As illustrated in Fig.~{\ref{fig:architecture}}, the proposed two-stage network, $\emph{i.e.,}$ TSCN, consists of two principal parts, namely CME-Net and CSR-Net. CME-Net uses the magnitude of the noisy spectrum as the input feature and the magnitude of the clean speech spectrum as the output, which is further coupled with the noisy phase to obtain a coarse complex spectrum (CCS), $\emph{i.e.,}$ $\left(\widetilde S^{cm}_{r}, \widetilde S^{cm}_{i} \right)$. In the second stage, both CCS and the original noisy spectrum are concatenated as the input of CSR-Net, and the network then estimates the residual spectrum, which is directly added to CCS to obtain a refined counterpart. Specifically, in the first stage, only the magnitude is optimized and most noise components can be removed. In the second stage, the network only needs to modify the phase while further refining the magnitude. In a nutshell, the calculation process is:
\begin{gather}
\label{eqn:equa1}
\lvert \tilde S^{cm} \rvert = \mathcal{F}_{cm}\left( \lvert X \rvert ;{\bf \Phi}_{cm} \right),\\
\left( \tilde S^{cs}_{r}, \tilde S^{cs}_{i} \right) = \left( \tilde S^{cm}_{r}, \tilde S^{cm}_{i} \right) + \mathcal{F}_{cs}\left(\tilde S^{cm}_{r}, \tilde S^{cm}_{i}, X_{r}, X_{i}; {\bf \Phi}_{cs} \right),
\end{gather}
where $\tilde S^{cm}_{r} = \Re \left(\lvert \tilde S^{cm} \rvert e^{j\theta_{X}} \right)$ and $\tilde S^{cm}_{i} = \Im \left(\lvert \tilde S^{cm} \rvert e^{j\theta_{X}} \right)$.

Both CME-Net and CSR-Net take the similar network topology as~{\cite{zhu2020flgcnn}}, which includes the gated convolutional encoder, decoder and stacked temporal convolution modules (dubbed TCMs)~{\cite{bai2018empirical}}. The encoder is utilized to extract the spectral patterns of the spectrum while the decoder is to reconstruct the spectrum. Note that instead of using long-short term memory (LSTM) as the basic unit for sequence modeling, stacked TCMs are utilized here to better capture both short and long range sequence dependencies.

\begin{figure}[t]
	\centering
	\centerline{\includegraphics[width=0.75\columnwidth]{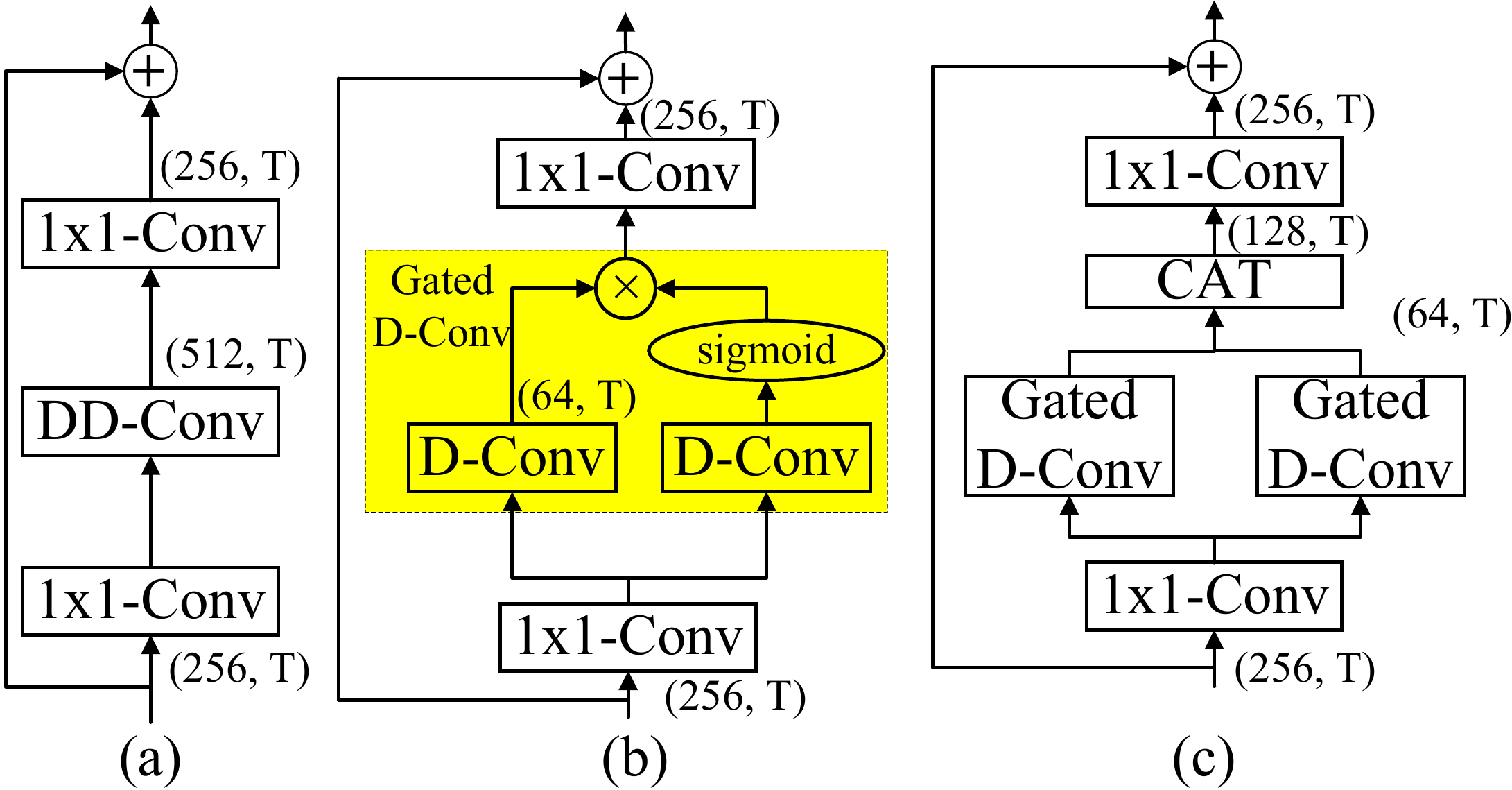}}
	\caption{Comparisons among different types of TCMs, both norm and activation layers are neglected for illustraction convenience. (a) Original TCM. (b) Proposed light-weight TCM. (c) Proposed light-weight dual TCM (DTCM).}
	\label{fig:tcm}
	\vspace{-0.6cm}
\end{figure}

As shown in Fig.~{\ref{fig:tcm}}(a), in the previous TCM settings, given input with size $\left(256, T\right)$ where $256$ and $T$ denote the number of channels and the timesteps, respectively, TCM first projects it to a higher channel space, $\emph{i.e.}$, 512, with input 1$\times$1-Conv, then dilated depthwise convolution (DD-Conv) is applied, the output 1$\times$1-Conv returns to 256. All the norm layers and activation functions are omitted for clarification convenience. To further lower the parameter burden, we propose two types of light-weight TCMs here, as shown in Fig.~{\ref{fig:tcm}}(b)-(c). In Fig.~{\ref{fig:tcm}}(b), the 1$\times$1-Conv compresses the channel number into 64, followed by gated D-Conv, $\emph{i.e.}$, a regular dilated convolution is multiplied with another dilated branch, where the sigmoid function is applied to scale the output values into $\left(0, 1\right)$. Fig.~{\ref{fig:tcm}}(c) is an improved version named DTCM, where two gated D-Convs are applied and the outputs from two branches are concatenated. Not that the dilation rates between two branches complement each other, $\emph{i.e.}$, if the dilation rate in one branch is $2^{r}$, then another dilation rate becomes $2^{M-r}$, where $M=5$ in this study. The rationale is that a large dilation rate means the long-range dependency can be captured while the local sequence correlation can be learned for a small dilation rate. As a result, the two branches establish both short and long sequence correlation during the training process. In this study, we adopt the TCM in Fig.~{\ref{fig:tcm}}(b) as the basic unit for CME-Net while the DTCM in Fig.~{\ref{fig:tcm}}(c) for CSR-Net.
\vspace{-0.3cm}
\subsection{Loss function}
\label{loss-function}
\vspace{-0.3cm}
For two-stage network, the following strategy is applied to train the network. First, we separately train the CME-Net with the following loss:
\begin{equation}
\label{eqn:equa2}
\mathcal{L}_{cm} = \left \| \lvert \tilde S^{cm} \rvert - \lvert S \rvert \right \|^{2}_{F},
\end{equation}
Afterward, the pretrained model of CME-Net is loaded and jointly optimized with CSR-Net, given as:
\begin{gather}
\label{eqn:equa3}
\mathcal{L} = \mathcal{L}_{cs}^{RI} + \mathcal{L}_{cs}^{Mag} + \lambda \mathcal{L}_{cm},\\
\mathcal{L}_{cs}^{RI} = \left \| \tilde{S}^{cs}_{r} - S_{r} \right \|^{2}_{F} + \left \| \tilde{S}^{cs}_{i} - S_{i} \right \|^{2}_{F},\\
\mathcal{L}_{cs}^{Mag} = \left \| \sqrt{ \lvert \tilde{S}_{r}^{cs}\rvert ^2 + \lvert \tilde{S}_{i}^{cs} \rvert ^2  } - \sqrt{ \lvert S_{r}\rvert ^2 + \lvert S_{i} \rvert ^2} \right \|_{F}^{2},
\end{gather}
where $\mathcal{L}_{cm}$, and $\mathcal{L}_{cs}^{*}$ denote the loss function of CME-Net and CSR-Net, respectively. $\lambda$ refers to the loss weight coefficient, which is set to 0.1 in this paper. Note that here two types of losses are considered for CSR-Net, namely RI loss and magnitude-based loss. The motivation can be explained from two aspects. Firstly, when the RI components are gradually optimized, the magnitude consistency cannot guarantee, $\emph{i.e.}$, the estimated magnitude may deviate its optimal optimization path~{\cite{wisdom2019differentiable}}. Secondly, experiments reveal that when magnitude constraint is set, consistent PESQ improvement can be achieved~{\cite{wang2020complex}}, which is helpful for speech quality.
\vspace{-0.2cm}
\subsection{Post-processing module}
\label{post-processing}
\vspace{-0.2cm}
With the MSE as the loss function, despite the notable improvement of speech quality, the residual noise components may become very unnatural, which may degrade the subjective quality. To improve the naturalness of the enhanced speech by deep learning approaches, many loss functions have been proposed, such as scale-invariant signal-to-noise ratio (SI-SDR)~{\cite{8683855}}, perceptual metric for speech quality evaluation (PMSQE)~{\cite{8468124}}, and the MSE with residual noise control~{\cite{li2020supervised}}. These loss functions can make the residual noise sound more natural than the MSE, and thus they can somewhat improve the speech quality.

In this paper, inspired by~{\cite{9054196, valin2018hybrid, 2013cepstrum}}, we use an extremely low-complexity deep learning approach similar to~{\cite{valin2018hybrid}} to further suppress the residual noise in the output of the pipeline 1. Instead of using the deep learning mapping gain directly on the estimated clean speech spectrum of the pipeline 1, we use this gain as an estimate of the speech presence probability (SPP) to recursively estimate the noise power spectral density (NPSD). With the estimated NPSD, the MMSE-LSA estimator is introduced to compute the final gain and then applies to suppress the residual noise. To further improve the robustness of the proposed PP scheme, we use a cepstrum-based preprocessing scheme to suppress the harmonic components before estimating the NPSD. By doing so, the over-estimation problem of the NPSD can be avoided in most cases.

\vspace{-0.2cm}

\vspace{-0.2cm}

\section{EXPERIMENTS}
\label{sec:experiments}
\vspace{-0.2cm}
\subsection{Datasets}
\label{datasets}
\vspace{-0.2cm}
In this study, we first explored the performance among different models on the WSJ0-SI84 dataset~{\cite{paul1992design}} to validate the performance superiority of the proposed two-stage network. Then the model, alongside with post-processing, was trained and evaluated with the ICASSP 2021 DNS-Challenge dataset to evaluate its performance on more complicated and real acoustic scenarios. For WSJ0-SI84, 5428, and 957 clean utterances with 77 speakers were selected to establish the datasets for training and validations, respectively. For the test set, 150 utterances are selected. Note that the speaker information in the test set is untrained. We randomly select 20,000 noises from DNS-Challenge{\footnote{https://github.com/microsoft/DNS-Challenge}} to form a 55 hours noise set for training. For testing, 4 challenging noises are selected, namely babble, cafe, and white from NOISEX92~{\cite{varga1993assessment}} and factory1 from CHIME3 dataset~{\cite{barker2015third}}. In this study, we create totally 50,000, 4000 noisy-clean pairs for training, and validation, respectively, with SNR ranging from -5\rm{dB} to 0\rm{dB}. The total duration for training is around 100 hours. 5 SNRs are selected for model evaluation, namely -5\rm{dB}, 0\rm{dB}, 5\rm{dB}, 10\rm{dB}, and 15\rm{dB}.

For the ICASSP 2021 DNS Challenge, relatively more complicated acoustic scenarios are considered than the INTERSPEECH 2020 DNS Challenge, including reverberation effect, cross-language, emotional, and singing cases. However, many provided utterances are relatively noisy, which are found to heavily impact the training convergence of the network. As a result, we drop the utterances with obviously poor quality. Totally, we generate a large noisy-clean training set with 517 hours duration, where about 65,000 provided noises are utilized and the SNR ranges from -5\rm{dB} to 25\rm{dB}. In addition, considering the reverberation effect in the real environment, around 30\% of utterances are convolved with 100,000 provided synthetic and real room impulse responses (RIRs) before mixed with different noise signals. The reverberation time T$_{60}$ ranges from 0.3 to 1.3 second in this study.
\vspace{-0.2cm}
\subsection{Parameter setup}
\label{parameter-setup}
\vspace{-0.2cm}
All the utterances are sampled at 16kHz. The 20ms Hanning window is adopted with 50\% overlap between consecutive frames. To extract the spectral features, 320-point FFT is utilized. Both models are optimized by Adam~{\cite{kingma2014adam}}. When the first model is separately trained, the initialized learning rate (LR) is set to 0.001. When the two models are jointly trained, LRs are set to 0.001 and 0.0001, respectively. The batch size is set to 8 at the utterance level. Note that to decrease the training time at the DNS-Challenge dataset, we directly finetune the models pre-trained on the WSJ0-SI84 to help the model adapt to the new dataset rapidly.
\vspace{-0.2cm}
\subsection{Compared models}
\label{compared-models}
\vspace{-0.2cm}
In this study, we compare the proposed two-stage network with another advanced 5 baselines, namely CRN~{\cite{tan2018convolutional}}, DARCN~{\cite{li2020recursive}}, TCNN~{\cite{pandey2019tcnn}}, GCRN~{\cite{tan2019learning}} and DCCRN~{\cite{hu2020dccrn}}, which are given as follows:

\begin{itemize}
	\item CRN: it is a causal convolutional recurrent network in the T-F domain. 5 convolutional and deconvolutional blocks are adopted as the encoder and decoder, respectively. 2 LSTM layers with 1024 units are adopted for sequence modeling. Only the magnitude is estimated while keeping the noisy phase unaltered. We keep the best configuration in~{\cite{tan2018convolutional}} and the number of parameters is 17.58M.
	
	\item DARCN: it is a causal convolutional network in the T-F domain, which combines recursive learning and dynamic attention mechanism together. The current estimation output is fed back to the input and the network is then reused to refine the estimation of the next stage. We keep the best configuration in~{\cite{li2020recursive}} and the number of parameters is 1.23M.
	
	\item TCNN: it is a causal encoder-TCMs-decoder topology defined in the time domain. Raw waveforms serve as both the input and output. We keep the same configuration in~{\cite{pandey2019tcnn}} and the number of parameters is 5.06M.
	
	\item GCRN: it is the advanced version of CRN, where both magnitude and phase are estimated. It has a similar topology with CRN except two decoders are utilized for RI estimation. We keep the best configuraion in~{\cite{tan2019learning}} and the number of parameters is 9.06M.
	
	\item DCCRN: it has achieved the first rank in the real-time track of Interspeech 2020 DNS-Challenge, where the complex-value operations are applied to CNN and RNN, and SI-SDR is utilized as the loss function. We keep the best configuration in~{\cite{hu2020dccrn}} and the number of parameters is 3.72M.
	
	\item TSCN: for the encoder and decoder parts, 5 (de)convolutional blocks are set, where the number of intermediate channels for each layer is 64, and the kernel size and stride are (2, 3) and (1, 2) in the time and frequency axis. For CME-Net, 18 light-weight TCMs are utilized for sequence learning, while 12 DTCMs are for CSR-Net{\footnote{In this paper, 6 (D)TCMs form a group, where the dilation rate in each group is (1, 2, 4, 8, 16, 32)}}. The number of parameter is 1.96M for CME-Net and 4.99M for TSCN.
\end{itemize}
\vspace{-0.4cm}
\section{RESULTS AND ANALYSIS}
\label{sec:results-and-analysis}
\vspace{-0.2cm}

\renewcommand\arraystretch{0.75}
\begin{table}[t]
	\caption{Objective results in terms of PESQ on the WSJ0-SI84 dataset. $\textbf{BOLD}$ denotes the best result in each case. ``Cau.'' denotes whether the system is causal implementation.}
	\centering
	\LARGE
	\resizebox{0.45\textwidth}{!}{
		\begin{tabular}{c|c|cccccc}
			\toprule
			Model &Cau. &-5\rm{dB} &0\rm{dB} &5\rm{dB} &10\rm{dB} &15\rm{dB} &Avg.\\
			\midrule
			Noisy &- &1.51 &1.84 &2.18 &2.54 &2.88 &2.19\\
			\midrule
			CRN &$\checkmark$ &1.95 &2.45 &2.83 &3.16 &3.44 &2.77\\
			DARCN &$\checkmark$ &2.04 &2.53 &2.89 &3.20 &3.47 &2.83 \\
			TCNN &$\checkmark$ &2.04 &2.53 &2.87 &3.17 &3.41 &2.80\\
			GCRN &$\checkmark$ &2.12 &2.64 &3.00 &3.29 &3.53 &2.92\\
			DCCRN &$\checkmark$ &2.08 &2.63 &3.03 &3.35 &3.63 &2.94\\
			CME-Net(Pro.) &$\checkmark$ &2.08 &2.55 &2.90 &3.23 &3.50 &2.85\\
			TSCN(Pro.) &$\checkmark$ &$\textbf{2.31}$ &$\textbf{2.83}$ &$\textbf{3.17}$ &$\textbf{3.44}$ &$\textbf{3.64}$ &$\textbf{3.08}$\\
			TSCN-PP(Pro.) &$\checkmark$ &2.07 &2.60 &2.91 &3.16 &3.37 &2.82 \\
			\bottomrule
	\end{tabular}}
	\label{tbl:pesq-comparison}
	\vspace*{-0.4cm}
\end{table}
\renewcommand\arraystretch{0.85}
\begin{table}[t]
	\caption{Objective results in terms of ESTOI (in \%) on the WSJ0-SI84 dataset. $\textbf{BOLD}$ denotes the best result in each case.}
	\centering
	\huge
	\resizebox{0.45\textwidth}{!}{
		\begin{tabular}{c|c|cccccc}
			\toprule
			Model &Cau. &-5\rm{dB} &0\rm{dB} &5\rm{dB} &10\rm{dB} &15\rm{dB} &Avg.\\
			\midrule
			Noisy &- &29.85 &44.92 &59.74 &73.49 &84.73 &58.55\\
			\midrule
			CRN &$\checkmark$ &48.80 &66.72 &78.23 &86.08 &91.30 &74.22\\
			DARCN &$\checkmark$ &50.92 &68.57 &79.11 &86.65 &91.74 &75.40 \\
			TCNN &$\checkmark$ &55.36 &72.44 &81.29 &87.27 &91.17 &77.50\\
			GCRN &$\checkmark$ &55.75 &72.61 &81.50 &87.68 &91.71 &77.85\\
			DCCRN &$\checkmark$ &56.56 &73.29 &83.13 &89.43 &$\textbf{93.43}$ &79.17\\
			CME-Net(Pro.) &$\checkmark$ &52.18 &68.80 &79.42 &86.81 &91.72 &75.79\\
			TSCN(Pro.) &$\checkmark$ &$\textbf{61.24}$ &$\textbf{76.49}$ &$\textbf{84.46}$ &$\textbf{89.57}$ &92.94 &$\textbf{80.94}$\\
			TSCN-PP(Pro.) &$\checkmark$ &60.73 &75.91 &83.90 &88.80 &92.09 &80.29 \\
			\bottomrule
	\end{tabular}}
	\label{tbl:estoi-comparison}
	\vspace*{-0.6cm}
\end{table}

\begin{figure}[t]
	\centering
	\centerline{\includegraphics[width=0.65\columnwidth]{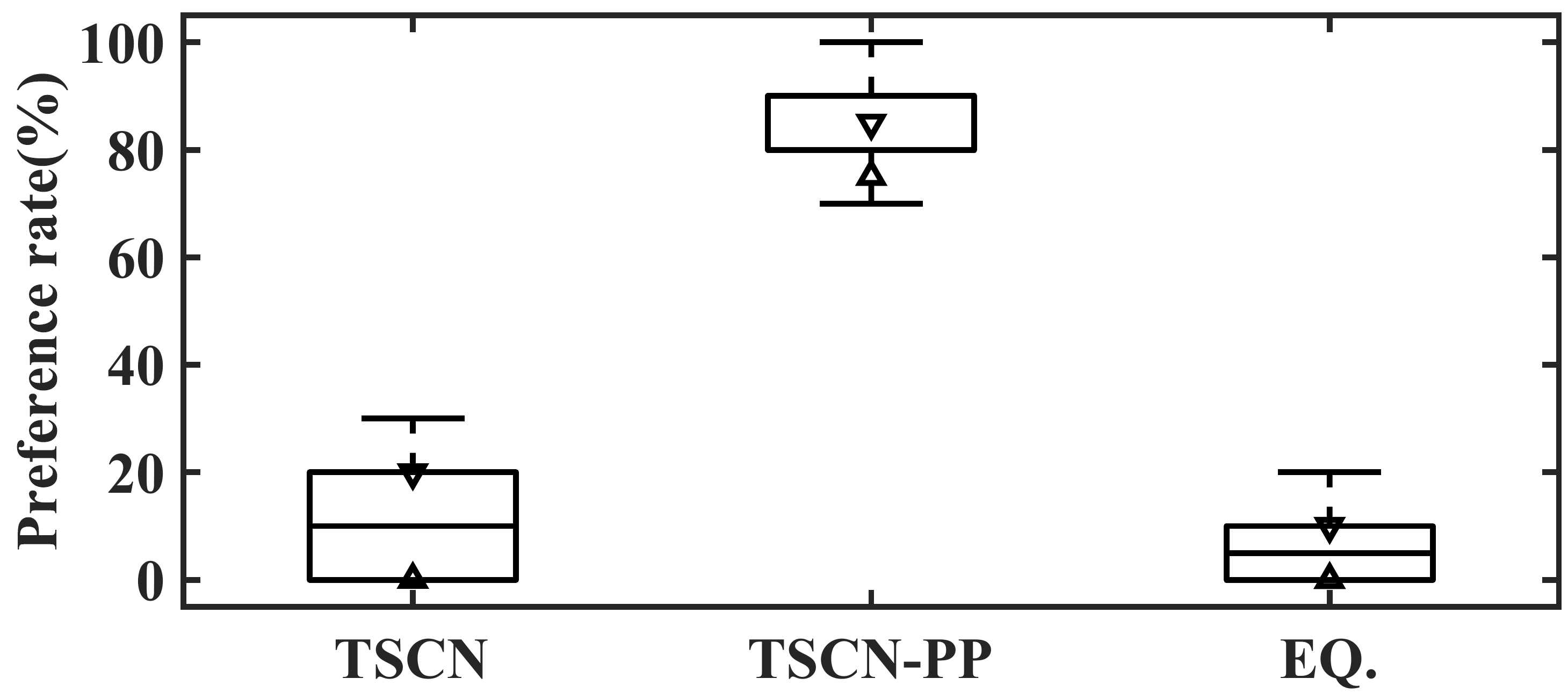}}
	\caption{Subjective evaluation test between TSCN and TSCN-PP. ``Equal" option is also provided if no decision can be made.}
	\label{fig:ab-test}
	\vspace*{-0.6cm}
\end{figure}
\vspace{-0.2cm}
\subsection{Objective comparisons}
\label{objective-comparisons}
\vspace{-0.2cm}
We utilize two objective metrics to evaluate the performance of different models, namely PESQ~{\cite{rix2001perceptual}} and ESTOI~{\cite{jensen2016algorithm}}, which are closely related to the human perceptual quality and intelligibility. The results are reported in Tables~{\ref{tbl:pesq-comparison}} and~{\ref{tbl:estoi-comparison}}. One can observe the following phenomena. Firstly, the proposed TSCN notably surpasses other baselines in both PESQ and ESTOI. For example, compared with DCCRN, a state-of-the-art approach, TSCN outperforms it by 0.14 and 1.77\% in terms of PESQ and ESTOI on average. This indicates the superior performance of the proposed method. Secondly, for relatively high SNRs, DCCRN seems to be relatively more advantageous. For example, in -5\rm{dB}, TSCN achieves about 0.23 PESQ improvement than DCCRN. However, for a high SNR like 15\rm{dB}, the PESQ scores are similar. Thirdly, when PP is applied, performance in both metrics are decreased. This is because PP is set here to suppress some unpleasant residual noise components, some speech components with low energy may also be cancelled. Nonetheless, we also believe that it is beneficial to implement PP as the network may generate some ``fake" spectral components under low SNR conditions, which sounds unpleasant. We expect that PP can effectively suppress the negative effects and improve the subjective quality, which will be validated in the next subsection. Overall, the proposed TSCN obtains impressive performance in objective metrics, motivating us to utilize it with PP for DNS-Challenge evaluation.
\vspace{-0.2cm}
\subsection{The impact of post-processing}
\label{impact-of-pp}
\vspace{-0.2cm}
To verify the impact of PP, we conduct the AB subjective test, which is similar to the procedure of~{\cite{li2020supervised}}. 10 volunteers are involved in the test. We randomly select 10 utterances from DNS blind test set, including 2 emotional, 3 English, 3 non-English, and 2 singing speech signals. Two processed types are provided, namely TSCN and TSCN-PP. The volunteers are required to select the preferred item with better subjective quality. ``Equal" option is also provided if no decision can be made. The testing results are illustrated in Fig.~{\ref{fig:ab-test}}. Compared with TSCN, after PP is applied, consistent subjective preference is achieved. This indicates the gap between objective metrics and subjective option, $\emph{i.e.}$, despite the notable degradation in the PESQ arises due to the spectrum information lost with PP, consistent subjective preference is still achieved as most unnatural residual noise is suppressed. It is interesting to see that this conclusion is consistent with the study in~{\cite{valin2020perceptually}}.
\renewcommand\arraystretch{1.25}
\begin{table}[t]
	\caption{Subjective evaluation with P.808 criterion on the DNS-Challenge.}
	\centering
	\Huge
	\resizebox{0.48\textwidth}{!}{
		\begin{tabular}{c|c|cccccc}
			\toprule
			Model &Track &Singing &Tonal &Non-English &English &Emotional &Overall\\
			\midrule
			Noisy &- &2.96 &3.00 &2.96 &2.80 &2.67 &2.86\\
			\midrule
			NSnet2(baseline) &RT &3.10 &3.25 &3.28 &3.30 &2.88 &3.21\\
			TSCN-PP(Pro.) &RT &$\textbf{3.14}$ &$\textbf{3.44}$ &$\textbf{3.50}$ &$\textbf{3.49}$ &$\textbf{2.92}$ &$\textbf{3.38}$\\
			\bottomrule
	\end{tabular}}
	\label{tbl:dns-challenge}
	\vspace*{-0.8cm}
\end{table}
\vspace{-0.2cm}
\subsection{Subjective results in the DNS-Challenge}
\label{subjective-results}
\vspace{-0.2cm}
In Table~{\ref{tbl:dns-challenge}}, we present the subjective results of the submission with ITU-T P.808 criterion~{\cite{reddy2020icassp}}, which is provided by the organizer. One can find that our method outperforms the baseline model by overall 0.17 MOS scores. Besides, the proposed method also achieves impressive performance on some unusual scenarios, like singing, tonal, and emotional, which are found considerably difficult to handle than conventional speech cases.

Finally, we evaluate the processing latency of the algorithm. In this study, the window size $T$ = 20ms, with overlap $T_{s}$ = 10ms between consecutive frames. As a result, the algorithmic delay $T_{d} = T + T_{s} = 30$ms, which meets the latency requirement. Note that no future information is utilized in this study, $i.e.,$ the system is strictly causal. We also calculate the processing time, and the average processing time per frame is 4.80ms for TSCN-PP and 3.84ms for TSCN tested on an Intel i5-4300U PC. Note that despite the two-stage network is applied, as we restrict the number of convolutional channels to 64 in each encoder and decoder, and meanwhile, LSTMs are replaced by parallelizable TCMs, the inference efficiency is still guaranteed.
\vspace{-0.3cm}
\section{CONCLUSIONS}
\label{sec:conclusion}
\vspace{-0.3cm}
In this challenge, we propose a novel system for denoising, which consists of a two-stage network and a low-complexity post-processing module. For the two-stage network, we decouple the optimization toward magnitude and phase, $i.e.$, the magnitude is first coarsely estimated, followed by the second network to refine phase information. To obtain better subjective quality, we also propose a light-weight post-processing module to further suppress the remaining unnatural residual noise, which usually arises when the test set mismatches the training conditions. Subjective results showed that the proposed algorithm ranked first in MOS for the real-time track 1 of the ICASSP 2021 DNS Challenge.



\vfill\pagebreak

\bibliographystyle{IEEEbib}
\bibliography{refs}

\begin{thebibliography}{10}

\bibitem{loizou2013speech}
Philipos~C Loizou,
\newblock {\em Speech enhancement: theory and practice},
\newblock CRC press, 2013.

\bibitem{xu2014regression}
Y.~Xu, J.~Du, L-R. Dai, and C-H. Lee,
\newblock ``A regression approach to speech enhancement based on deep neural
  networks,''
\newblock {\em IEEE/ACM Trans.~Audio Speech Lang.~Proc.}, vol. 23, no. 1, pp.
  7--19, 2014.

\bibitem{wang2018supervised}
D.~Wang and J.~Chen,
\newblock ``Supervised speech separation based on deep learning: An overview,''
\newblock {\em IEEE/ACM Trans.~Audio Speech Lang.~Proc.}, vol. 26, no. 10, pp.
  1702--1726, 2018.

\bibitem{wang1982unimportance}
D.~Wang and J.~Lim,
\newblock ``The unimportance of phase in speech enhancement,''
\newblock {\em IEEE Transactions on Acoustics, Speech, and Signal Processing},
  vol. 30, no. 4, pp. 679--681, 1982.

\bibitem{paliwal2011importance}
K.~Paliwal, K.~W{\'o}jcicki, and B.~Shannon,
\newblock ``The importance of phase in speech enhancement,''
\newblock {\em Speech Commun.}, vol. 53, no. 4, pp. 465--494, 2011.

\bibitem{williamson2017time}
D.~Williamson and D.~Wang,
\newblock ``Time-frequency masking in the complex domain for speech
  dereverberation and denoising,''
\newblock {\em IEEE/ACM Trans.~Audio Speech Lang.~Proc.}, vol. 25, no. 7, pp.
  1492--1501, 2017.

\bibitem{tan2019learning}
K.~Tan and D.~Wang,
\newblock ``Learning complex spectral mapping with gated convolutional
  recurrent networks for monaural speech enhancement,''
\newblock {\em IEEE/ACM Trans.~Audio Speech Lang.~Proc.}, vol. 28, pp.
  380--390, 2020.

\bibitem{defossez2020real}
A.~D{\'e}fossez, G.~Synnaeve, and Y.~Adi,
\newblock ``Real {T}ime {S}peech {E}nhancement in the {W}aveform {D}omain,''
\newblock {\em in Proc. Interspeech 2020}, pp. 3291--3295, 2020.

\bibitem{reddy2020icassp}
C.~Reddy, H.~Dubey, V.~Gopal, R.~Cutler, S.~Braun, H.~Gamper, R.~Aichner, and
  S.~Srinivasan,
\newblock ``{ICASSP} 2021 {D}eep {N}oise {S}uppression {C}hallenge,''
\newblock {\em arXiv preprint arXiv:2009.06122}, 2020.

\bibitem{yu2019gradual}
W.~Yu, Z.~Huang, W.~Zhang, L.~Feng, and N.~Xiao,
\newblock ``Gradual network for single image de-raining,''
\newblock in {\em Proc. of ACMM}, 2019, pp. 1795--1804.

\bibitem{li2020speech}
A.~Li, M.~Yuan, C.~Zheng, and X.~Li,
\newblock ``Speech enhancement using progressive learning-based convolutional
  recurrent neural network,''
\newblock {\em Appl. Acoust.}, vol. 166, pp. 107347, 2020.

\bibitem{li2020recursive}
A.~Li, C.~Zheng, C.~Fan, R.~Peng, and X.~Li,
\newblock ``A recursive network with dynamic attention for monaural speech
  enhancement,''
\newblock {\em in Proc. of Interspeech 2020}, 2020.

\bibitem{zhu2020flgcnn}
Y.~Zhu, X.~Xu, and Z.~Ye,
\newblock ``{FLGCNN}: A novel fully convolutional neural network for end-to-end
  monaural speech enhancement with utterance-based objective functions,''
\newblock {\em Appl. Acoust.}, vol. 170, pp. 107511, 2020.

\bibitem{bai2018empirical}
S.~Bai, J.~Kolter, and V.~Koltun,
\newblock ``An empirical evaluation of generic convolutional and recurrent
  networks for sequence modeling,''
\newblock {\em arXiv preprint arXiv:1803.01271}, 2018.

\bibitem{wisdom2019differentiable}
S.~Wisdom, J.~Hershey, K.~Wilson, J.~Thorpe, M.~Chinen, B.~Patton, and
  R.~Saurous,
\newblock ``Differentiable consistency constraints for improved deep speech
  enhancement,''
\newblock in {\em Proc. of ICASSP}. IEEE, 2019, pp. 900--904.

\bibitem{wang2020complex}
Z.~Wang, P.~Wang, and D.~Wang,
\newblock ``Complex spectral mapping for single-and multi-channel speech
  enhancement and robust {ASR},''
\newblock {\em IEEE/ACM Trans.~Audio Speech Lang.~Proc.}, vol. 28, pp.
  1778--1787, 2020.

\bibitem{8683855}
J.~L. {Roux}, S.~{Wisdom}, H.~{Erdogan}, and J.~R. {Hershey},
\newblock ``{SDR} – {H}alf-baked or well done?,''
\newblock in {\em Proc. of ICASSP}, 2019, pp. 626--630.

\bibitem{8468124}
J.~M. {Martin-Doñas}, A.~M. {Gomez}, J.~A. {Gonzalez}, and A.~M. {Peinado},
\newblock ``A deep learning loss function based on the perceptual evaluation of
  the speech quality,''
\newblock {\em IEEE Signal Process. Lett.}, vol. 25, no. 11, pp. 1680--1684,
  2018.

\bibitem{li2020supervised}
A.~Li, R.~Peng, C.~Zheng, and X.~Li,
\newblock ``A supervised speech enhancement approach with residual noise
  control for voice communication,''
\newblock {\em Appl. Sci.}, vol. 10, no. 8, pp. 2894, 2020.

\bibitem{9054196}
M.~{Tammen}, D.~{Fischer}, B.~T. {Meyer}, and S.~{Doclo},
\newblock ``{DNN}-based speech presence probability estimation for multi-frame
  single-microphone speech enhancement,''
\newblock in {\em Proc. of ICASSP}, 2020, pp. 191--195.

\bibitem{valin2018hybrid}
Jean-Marc Valin,
\newblock ``A hybrid {DSP}/deep learning approach to real-time full-band speech
  enhancement,''
\newblock in {\em Proc. of MMSP}. IEEE, 2018, pp. 1--5.

\bibitem{2013cepstrum}
X.~Hu, S.~Wang, C.~Zheng, and X.~Li,
\newblock ``A cepstrum-based preprocessing and postprocessing for speech
  enhancement in adverse environments,''
\newblock {\em Appl. Acoust.}, vol. 74, no. 12, pp. 1458--1462, 2013.

\bibitem{paul1992design}
D.~Paul and J.~Baker,
\newblock ``The design for the wall street journal-based {CSR} corpus,''
\newblock in {\em Workshop on Speech and Natural Language}, 1992, p. 357–362.

\bibitem{varga1993assessment}
A.~Varga and H.~Steeneken,
\newblock ``Assessment for automatic speech recognition: {II}. {NOISEX}-92: A
  database and an experiment to study the effect of additive noise on speech
  recognition systems,''
\newblock {\em Speech Commun.}, vol. 12, no. 3, pp. 247--251, 1993.

\bibitem{barker2015third}
J.~Barker, R.~Marxer, E.~Vincent, and S.~Watanabe,
\newblock ``The third ‘chime’speech separation and recognition challenge:
  Dataset, task and baselines,''
\newblock in {\em Proc. of ASRU}. IEEE, 2015, pp. 504--511.

\bibitem{kingma2014adam}
D.~Kingma and J.~Ba,
\newblock ``Adam: A method for stochastic optimization,''
\newblock {\em arXiv preprint arXiv:1412.6980}, 2014.

\bibitem{tan2018convolutional}
K.~Tan and D.~Wang,
\newblock ``A convolutional recurrent neural network for real-time speech
  enhancement.,''
\newblock in {\em Proc. of Interspeech}, 2018, pp. 3229--3233.

\bibitem{pandey2019tcnn}
A.~Pandey and D.~Wang,
\newblock ``T{CNN}: Temporal convolutional neural network for real-time speech
  enhancement in the time domain,''
\newblock in {\em Proc. of ICASSP}. IEEE, 2019, pp. 6875--6879.

\bibitem{hu2020dccrn}
Y.~Hu, Y.~Liu, S.~Lv, M.~Xing, S.~Zhang, Y.~Fu, J.~Wu, B.~Zhang, and L.~Xie,
\newblock ``{DCCRN}: Deep complex convolution recurrent network for phase-aware
  speech enhancement,''
\newblock in {\em Proc. of Interspeech 2020}, 2020, pp. 2472--2476.

\bibitem{rix2001perceptual}
A.~Rix, J.~Beerends, M.~Hollier, and A.~Hekstra,
\newblock ``Perceptual evaluation of speech quality ({PESQ})-a new method for
  speech quality assessment of telephone networks and codecs,''
\newblock in {\em Proc. of ICASSP}. IEEE, 2001, vol.~2, pp. 749--752.

\bibitem{jensen2016algorithm}
J.~Jensen and C.~Taal,
\newblock ``An algorithm for predicting the intelligibility of speech masked by
  modulated noise maskers,''
\newblock {\em IEEE/ACM Trans.~Audio Speech Lang.~Proc.}, vol. 24, no. 11, pp.
  2009--2022, 2016.

\bibitem{valin2020perceptually}
J.~Valin, U.~Isik, N.~Phansalkar, R.~Giri, K.~Helwani, and A.~Krishnaswamy,
\newblock ``A perceptually-motivated approach for low-complexity, real-time
  enhancement of fullband speech,''
\newblock {\em in Proc. of Interspeech 2020}, 2020.

\end{thebibliography}

\end{document}